\documentclass[twocolumn,twoside]{article}
\pdfoutput=1
\usepackage{authblk}
\usepackage[]{gensymb}
\usepackage[]{graphicx}
\usepackage[detect-weight=true, detect-family=true]{siunitx}
\DeclareSIUnit\torr{Torr}
\usepackage[caption=false]{subfig}

\newcommand{\uv}{\SI{243}{\nm} } 
\newcommand{\ir}{\SI{972}{\nm} } 
\newcommand{\blue}{\SI{486}{\nm} } 
\title{Cavity enhanced DUV laser for two-photon cooling of atomic hydrogen}

\author[1,*]{S. F. Cooper}
\author[1]{Z. Burkley}
\author[1]{A. D. Brandt}
\author[1]{C. Rasor}
\author[1]{D. C. Yost}

\affil[1]{Physics Department, Colorado State University, 1875 Campus Delivery, Fort Collins, CO, 80521}

\affil[*]{Corresponding author: Samuel.Cooper@rams.colostate.edu}

\begin{document}

\maketitle

\begin{abstract}
{We demonstrate a \SI{650}{\mW} \SI{243}{\nm} continuous-wave laser coupled to a linear optical enhancement cavity. The enhancement cavity can maintain $>$\SI{30}{\W} of intracavity power for \SI{1}{\hour} of continuous operation without degradation. This system has sufficient power for a demonstration of two-photon laser cooling of hydrogen and may be useful for experiments on other simple two-body atomic systems.}
\end{abstract}

\section{Introduction}

Radiation with a wavelength of 243 nm is required to excite the 1S-2S two-photon transition in atomic hydrogen. This transition possesses a narrow width of only \SI{1.3}{\Hz} and is particularly well-suited for precision measurement. After decades of improvement \cite{ hansch:1975, foot:1985, boshier:1989, parthey:2011}, the 1S-2S energy interval is now known with an incredibly small fractional uncertainty of $4.2 \times 10^{-15}$ \cite{parthey:2011}.  In addition, optical excitation of this transition allows for subsequent excitation to higher states. Spectroscopy of these other transitions in hydrogen, along with the 1S-2S, collectively determine the Rydberg constant, the radius of the proton, and ultimately provide a test of quantum electrodynamics \cite{Beauvoir:2000,codata:2014,beyer:2017}. Currently, the proton charge radius as determined through spectroscopy of normal hydrogen is discrepant with the value determined through spectroscopy of muonic hydrogen \cite{codata:2014}. This highlights the need for additional hydrogen spectroscopy using new techniques which aim to reduce systematic effects.

Two-photon excitation to the 2S state requires high power, narrow bandwidth radiation, and developing these sources has been actively pursued for many decades \cite{couillaud:1984,zimmermann:1995,kolachevsky:2006}.  While \numrange{50}{100} \si{\milli\W} can now be routinely produced \cite{beyer:2013}, for experiments on hydrogen and other two-body atomic systems, there are many motivations for continuing to power scale these lasers. For example, when performing 1S-2S spectroscopy, additional laser power would allow the beam to be transversely expanded to increase the interaction volume. For normal hydrogen, a larger interaction volume would decrease the time-of-fight broadening and increase the fraction of the atomic beam excited, reducing the statistical uncertainty \cite{parthey:2011}.  Increasing the interaction volume is also pertinent for 1S-2S spectroscopy in antihydrogen. The excitation of this transition took $\approx$ 300 seconds per data point \cite{ahmadi:2017} due to the relatively large trap volume as compared to the interaction volume. Finally, if 243 nm lasers can be power scaled sufficiently, it may be possible to laser cool hydrogen using the 1S-2S transition \cite{zehnle:2001, kielpinski:2006,Wu:2011}. Such two-photon laser cooling is the primary motivation for the work presented in this letter. 

In spite of significant interest, hydrogen laser cooling has proven notoriously difficult. While single photon cooling using the 1S-2P transition is conceptually straightforward, it requires Lyman-$\alpha$ radiation at 121.6 nm which is difficult to produce and manipulate using standard optical components. For example, there are no nonlinear crystals without significant losses so frequency conversion is typically performed in gases (see for example \cite{hilbig:1981,Eikema:2001} ).  In addition, MgF$_2$ lenses used for collimation will only transmit between 50-60\% of the power in the best case, and propagation must be performed in vacuum to avoid absorption in air.  Due to these difficulties, there has been only one demonstration of laser cooling using Lyman-$\alpha$ radiation \cite{setija:1993}. The laser cooling was performed within a cryogenic magnetic trap and it took several minutes to complete due to the low Lyman-$\alpha$ power ($\approx$ 2 nW) at the atomic sample. Therefore, even though a majority of systematic uncertainties in hydrogen spectroscopy can be traced back to the finite temperature of the atoms, most measurements have been limited to atomic beams with a temperature no lower than \SI{4}{\K} -- orders of magnitude larger than laser cooled atom temperatures \cite{parthey:2011,beyer:2017,Beauvoir:2000}. 

To avoid some of the difficulties which are inherent to Lyman-$\alpha$ laser sources, Zehnle {\it et. al} proposed using the 1S-2S two-photon transition itself for laser cooling \cite{zehnle:2001}.  In contrast to single-photon cooling, this requires radiation at 243 nm, which can be produced and manipulated much more easily. One complication is that, since the 2S state is metastable, the population must be quenched by mixing with the 2P \cite{zehnle:2001,kielpinski:2006} or 3P  state \cite{Wu:2011}. However, a more significant challenge is posed by the average power required. While cavity enhanced 243 nm lasers have previously reached $\approx 1$ \si{\W} \cite{beyer:2013}, useful scattering rates for laser cooling would require average power at least an order of magnitude higher \cite{kielpinski:2006}. To the best of our knowledge, this level of intracavity power has not been previously demonstrated in the deep ultraviolet (DUV: 200 nm - 280 nm), and the conditions, if any, under which such a cavity can operate has not been previously investigated.  

There are a number of difficulties associated with the operation of an enhancement cavity in the deep ultraviolet and with high intracavity power. First, commercially available mirror coatings are only able to achieve reflectivity of about \SI{99.5}{\%}, and for a perfectly impedance matched cavity with no other losses, the maximum enhancement is limited to $\approx \, 100$. Therefore, the only clear avenue to increasing the intracavity power is to power scale the excitation laser. This was pursued in our previous work, where we demonstrated cw outputs of $\approx$ \SI{600}{\mW} at 243 nm \cite{burkley:2016}.  Second, short wavelength optics are known to degrade when exposed to high power due to surface oxygen depletion \cite{Gangloff:15} and hydrocarbon contamination \cite{Kunz:2000}. In the UV \cite{Gangloff:15}, and extreme ultraviolet \cite{Heinbuch:08} degradation due to both effects has been mitigated by admitting O$_2$ at the location of the mirrors -- albeit at a lower average power than what is required for two-photon laser cooling. 

Here, we demonstrate cavity enhanced \uv power of $>$\SI{30}{\W} utilizing a high power excitation laser. By flushing the cavity mirrors with O$_2$ and utilizing differential pumping, we could stably lock the cavity for more than one hour without degradation.  As described in \cite{kielpinski:2006}, this level of intracavity power should be sufficient for an initial demonstration of laser cooling. 

\section{Experiment}


\begin{figure}[tb]
\includegraphics[clip,width=\linewidth]{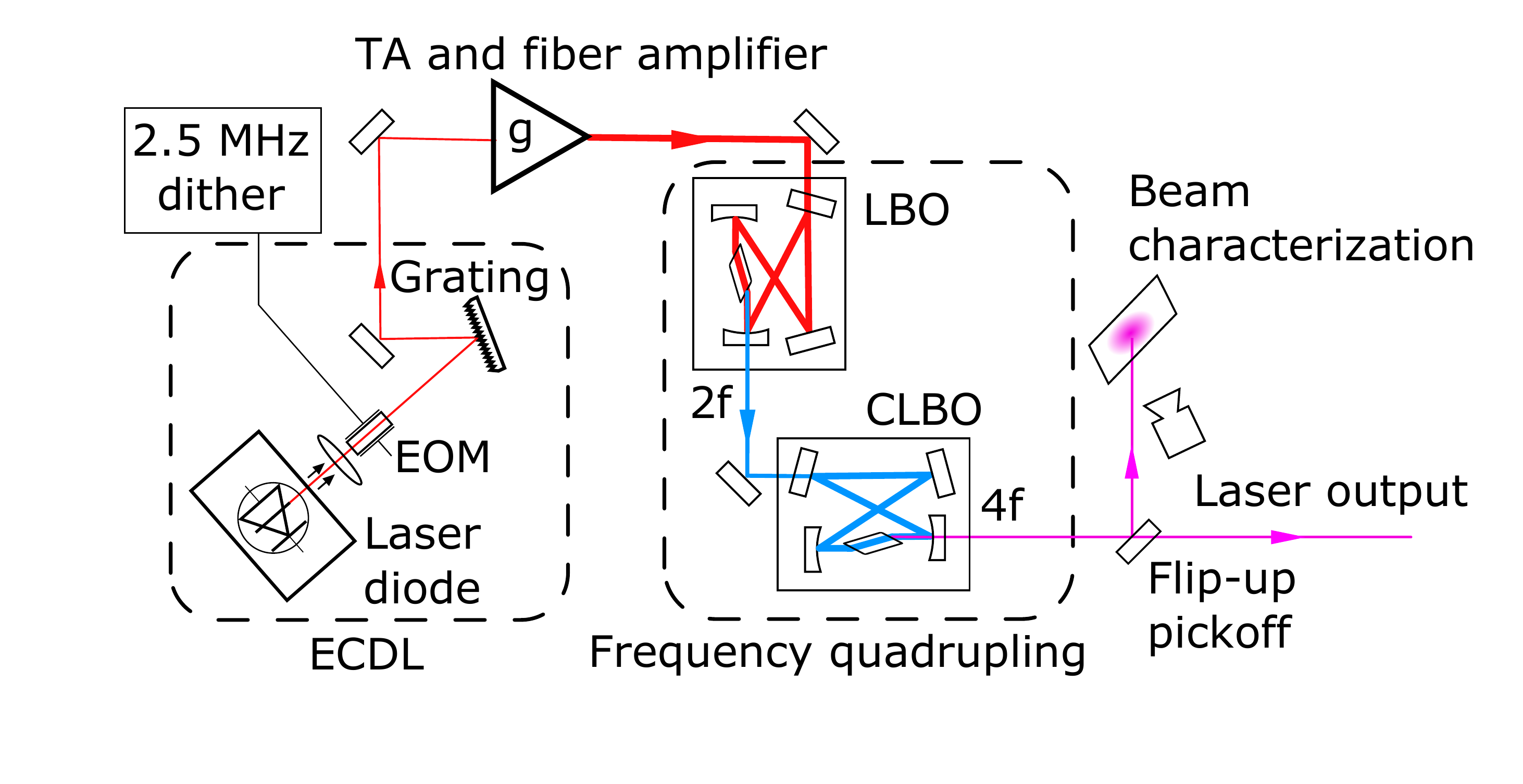}%
\caption{The master oscillator is an extended cavity diode laser (ECDL) at \SI{972}{\nm} and amplified as in \cite{burkley:2016}.  An EOM within the cavity applies a \SI{2.5}{\MHz} dither for locking the three subsequent enhancement cavities. The IR light is frequency quadrupled with two bow-tie enhancement cavities -- first from \ir to \blue using LBO with noncritical (temperature) phase matching, then from \blue to \uv using CLBO with critical phase matching. We characterize the \uv output with a fluorescent screen and CCD camera.}
\label{fig:laser_schematic}
\end{figure}


\begin{figure}[htbp]
\includegraphics[clip,width=\linewidth]{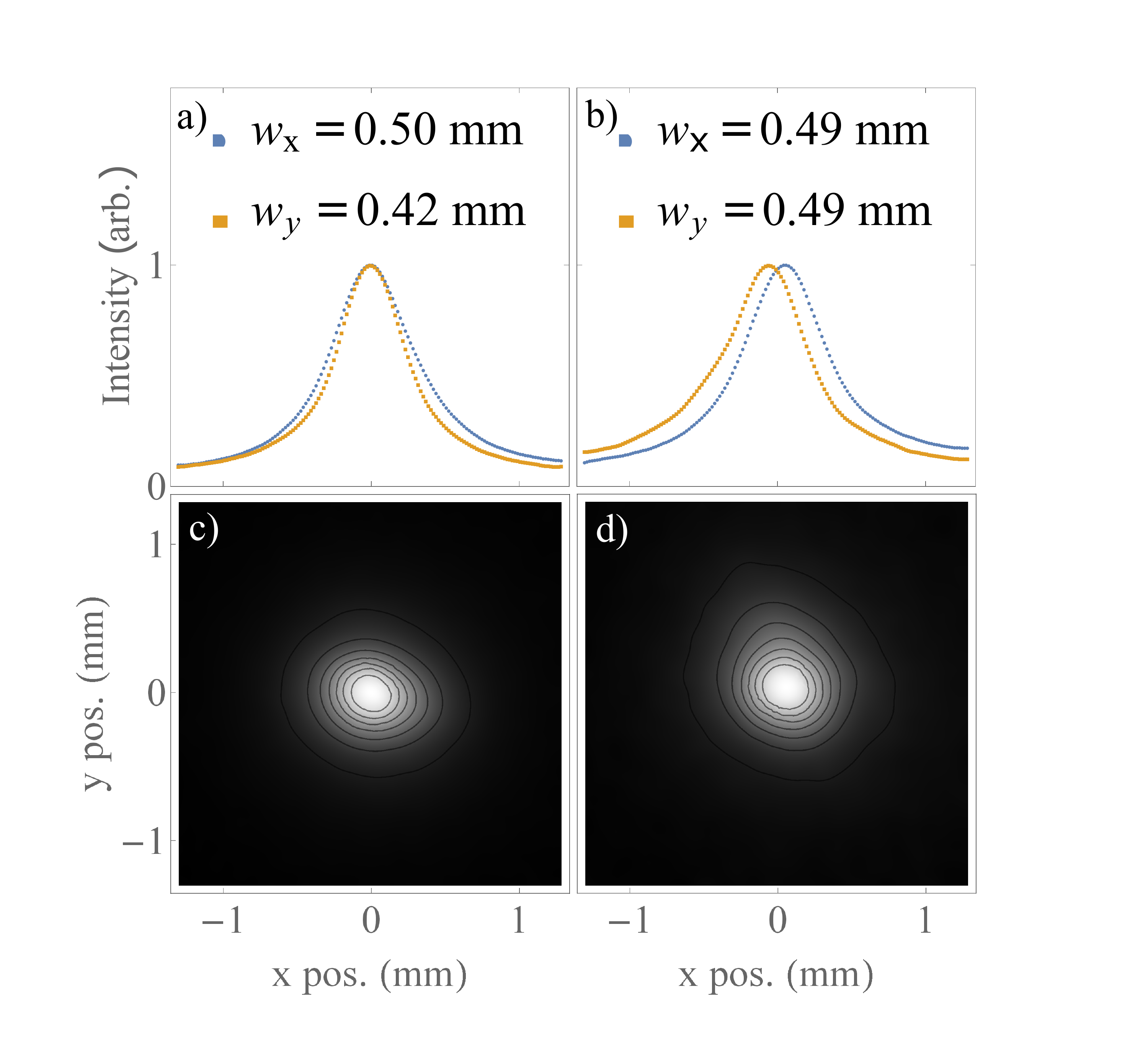}%
\caption{a) Profile of the laser output directly after the telescoping lenses, b) profiles of the laser output \SI{114}{\cm} after the lenses. c) Image corresponding to the profiles in a). d) Image corresponding to the profiles in b). The beam images are taken at a power of \SI{386}{\mW} and $w_x$ and $w_y$ are respectively the x and y $1/e^2$ waists from fits to the intensity data.
}
\label{fig:cavity_coupling}
\end{figure}

The 243 nm laser used for the excitation of our enhancement cavity was previously described in \cite{burkley:2016}, and an abbreviated schematic is shown in Fig. \ref{fig:laser_schematic}. First, a \SI{972}{\nm} laser diode seed is amplified by a commercial tapered amplifier (TA) and a Yb-fiber amplifier. The Yb-fiber amplifier allowed for power scaling and operated at 972 nm where the system behaves effectively as a 3-level system \cite{Nilsson:1998}. After amplification, the \SI{972}{\nm} light is frequency quadrupled to \SI{243}{\nm} using two successive resonant doubling stages.  For the first doubling stage, we use an LBO crystal in a noncritical phase matching configuration which produces a symmetric 486 nm beam profile due to the absence of walkoff effects. For the second doubling stage, we use a CLBO crystal in a critical phase matching configuration, which produces an elliptic beam with an aspect ratio of roughly three-to-one. This source is able to consistently produce $\approx$ \SI{650}{\mW} of \uv power. With careful alignment, we have observed $>1$ W of 243 nm power for a few minutes although it is difficult to maintain the higher power over hour time scales without realignment of the system. We attribute this behavior to thermal instability since we have observed no degradation of the CLBO crystal -- even under the highest power conditions.

\begin{figure}[t]
\includegraphics[clip,width=\linewidth]{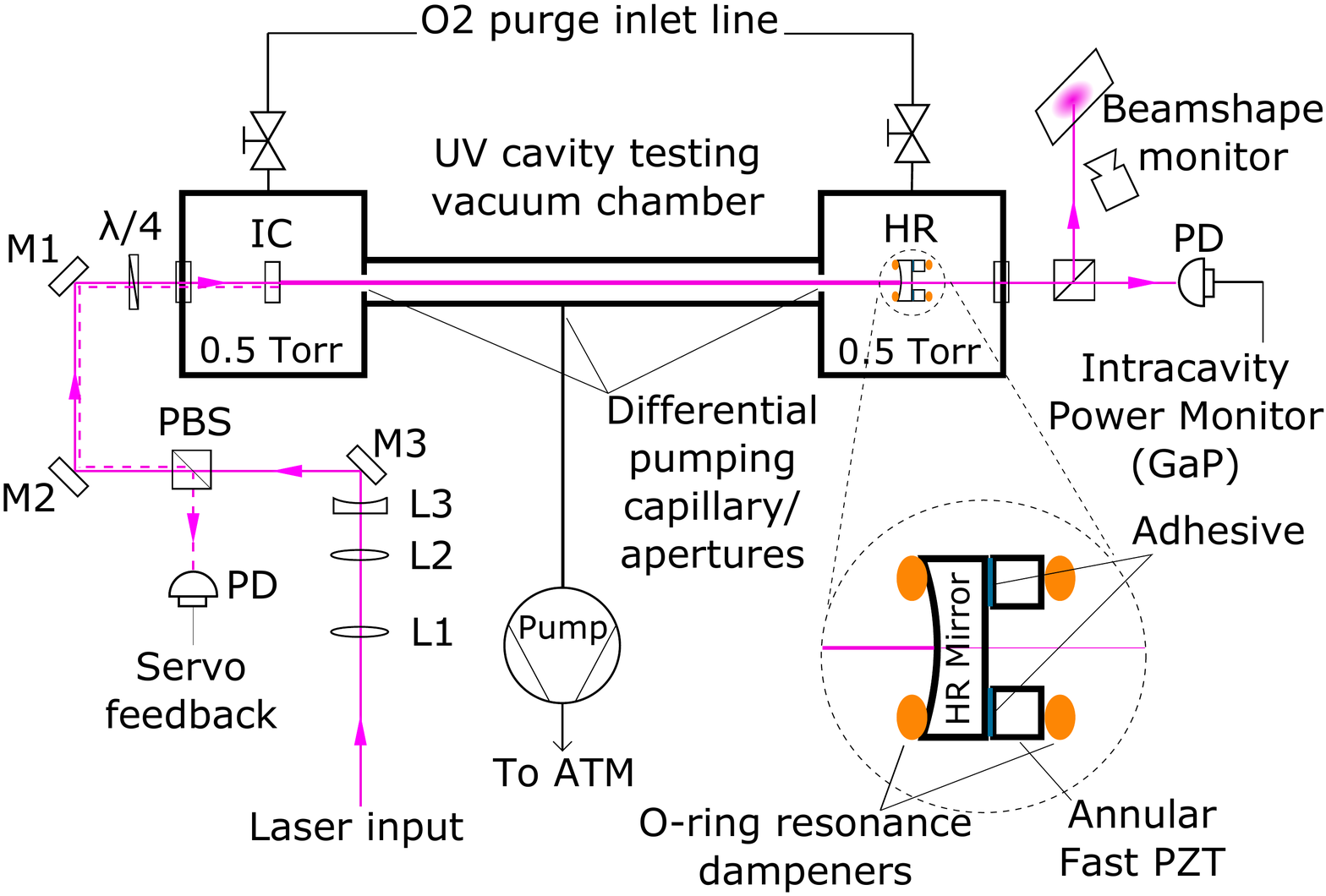}%
\caption{Diagram of enhancement cavity. The input \uv beam is mode matched to the cavity using one spherical lens (L1) and two cylindrical lenses (L2 and L3). The reflection from the input coupler (IC) is optically isolated from the laser with a polarizing beam splitter (PBS) and quarter waveplate ($\lambda/4$). The transmission of the high reflector (HR) is monitored with a photodiode (PD) to determine the intracavity power.}
\label{fig:apparatus_schematic}
\end{figure}

Mode-matching to the enhancement cavity is accomplished with two cylindrical lenses and one spherical lens. Since the walkoff in the last doubling cavity produces an elliptic profile, cylindrical lenses are necessary to reshape the beam.  Figure \ref{fig:cavity_coupling} shows the beam profile in two different locations separated by \SI{1.14}{\m} after reshaping. As can be seen, the beam has a nearly symmetric, Gaussian intensity profile.
As shown in Fig. \ref{fig:apparatus_schematic} the mode-matching optics are followed by a polarizing beam splitter and $\lambda$/4 waveplate, which are used to isolate the cavity from the second doubling stage.  In general, transmission optics in the DUV suffer from high loss. Each optic used for mode matching and beam isolation have losses ranging from 1\% to 8\%. This results in a maximum stable power at the cavity input of \SI{420}{\mW}.  

The enhancement cavity is formed with a high reflector (HR) with a radius of curvature $R = \SI{4}{\m}$ and a flat input coupler (IC) spaced by 0.75 m (free spectral range, FSR = 200 MHz). The resonant TEM$_{00}$ mode of the cavity has a focus at the input coupler with a calculated beam waist of $w_0=350$ $\mu$m. The calculated waist at the curved high reflector is $w_0=390$ $\mu$m so that the best mode-matching is achieved with a nearly collimated input beam. The overlap integral of the beam defined by the profiles in Fig. \ref{fig:cavity_coupling} with the enhancement cavity TEM$_{00}$ mode is \SI{\approx ~ 92}{\%}. 

\begin{figure}[t]
\includegraphics[clip,width=\linewidth]{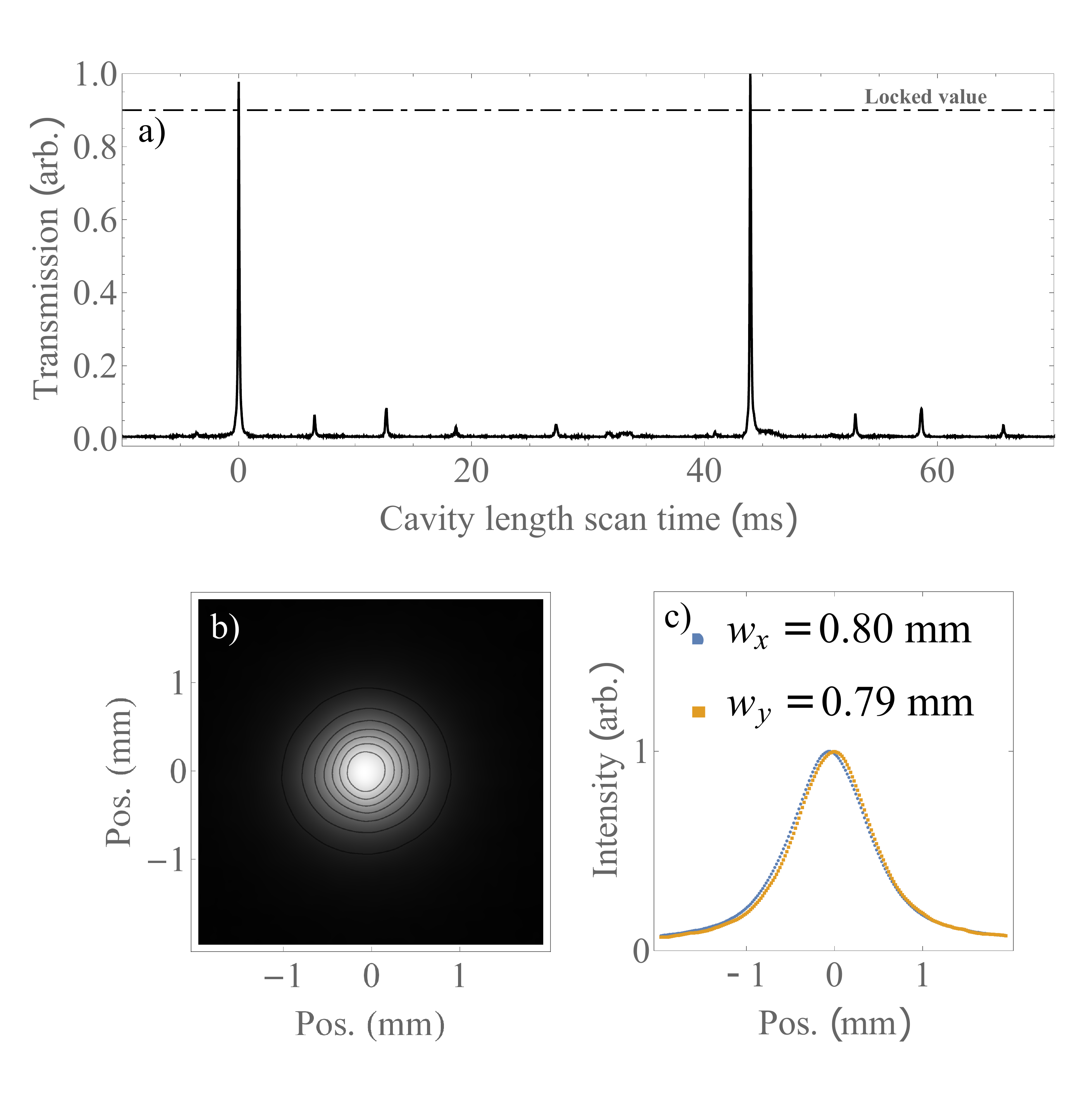}%
\caption{a) Cavity transmission as a function of frequency. The resonance width allows us to estimate the finesse to be $\approx 300$ with $>\SI{80}{\%}$ of the power in the TEM$_{00}$ mode. b) Image of cavity transmission with \SI{31}{\W} of intracavity power.  The image was taken \SI{65.5}{\cm} after the output coupler with a build up of \num{75}. c) Cavity transmission profile.}
\label{fig:mode_matching}
\end{figure}

The dielectric cavity mirrors were designed for high power and reflectivity using electron beam evaporation (LaserOptik GmbH).  The mirrors have a SiO$_2$ top surface layer which according to \cite{Gangloff:15} is helpful to reduce oxygen depletion of the coating. Since we expected combined absorption and scattering losses of 0.5\%  for both cavity optics, we chose the input coupler to have a 1.5\% transmission so that the cavity is slightly overcoupled. Figure \ref{fig:mode_matching} shows the cavity transmission as the length is scanned. From the ratio of the resonance widths to the FSR, we estimate that the finesse is $\approx 300$. 


We have carefully calibrated the transmission of the high reflector. As shown in Fig. \ref{fig:apparatus_schematic}, we monitor the transmission with a GaP photodiode to determine the intracavity power.  
When scanning over the resonance, the intracavity power reaches \SI{38}{\W} with an input power of 420 mW.  This corresponds to a buildup of $\approx 92$.  As can be seen in Fig.\ref{fig:mode_matching}, there is minimal excitation of higher-order cavity modes due to imperfect mode-matching and more than 80\% of the power is within the TEM$_{00}$ mode. When taking into account the buildup, finesse and mode-matching factor, we estimate that the total scattering and absorption loss of the cavity is $\approx 0.8$\%. This is slightly better than the 1\% estimate from the mirror manufacturer. As the power of the laser is increased, we observed that the vertical waist of the beam at the output of the final doubling stage increases slightly. We attribute this to a thermal variation in the CLBO crystal. To compensate for this, we readjusted the cylindrical lenses used to mode-match to the cavity as the laser power was varied. 

The enhancement cavity is kept on resonance using a Pound-Drever-Hall (PDH) locking scheme \cite{drever:1983}. For this, we generate sidebands on the fundamental radiation at 972 nm using an electro-optic modulator (EOM) within the master oscillator. The sidebands are also imprinted on the quadrupled radiation which are used to generate a PDH error signal for the DUV enhancement cavity.  As shown in Fig. \ref{fig:apparatus_schematic}, the cavity is locked to the laser utilizing a fast piezo-electric transducer (PZT). The fast PZT is annular so that we can measure the light transmitted through the high reflecting mirror. We use the mirror mounting procedure in \cite{Goldovsky:16} to damp low frequency resonances, which is crucial to maintain a stable lock. When locked, we measure a peak intracavity power of \SI{33.7}{\W} circulating with a build up 80.4 and an input of \SI{420}{\mW}. The PZT mounted  cavity mirror is itself on an optical stage whose position is adjusted with a large throw PZT, which compensates for slow drifts. 

When compared with the maximum buildup as the cavity is scanned over the resonance, the locked intracavity power is consistently $\SI{10}{\%}$ lower (as indicated in Fig. \ref{fig:mode_matching}a). The power decays to this steady state value within \SI{40}{\ms} after the lock is engaged.  Due to the time scale, we attribute this behavior to heating of the cavity mirror coatings.  However, we monitored the transmitted beam profile simultaneously (as shown in Fig. \ref{fig:mode_matching}b--c) and we observed that the beam shape was nearly Gaussian and that its profile did not change noticeably with power.

\begin{figure}[tb]
\centering
\includegraphics[width=\linewidth]{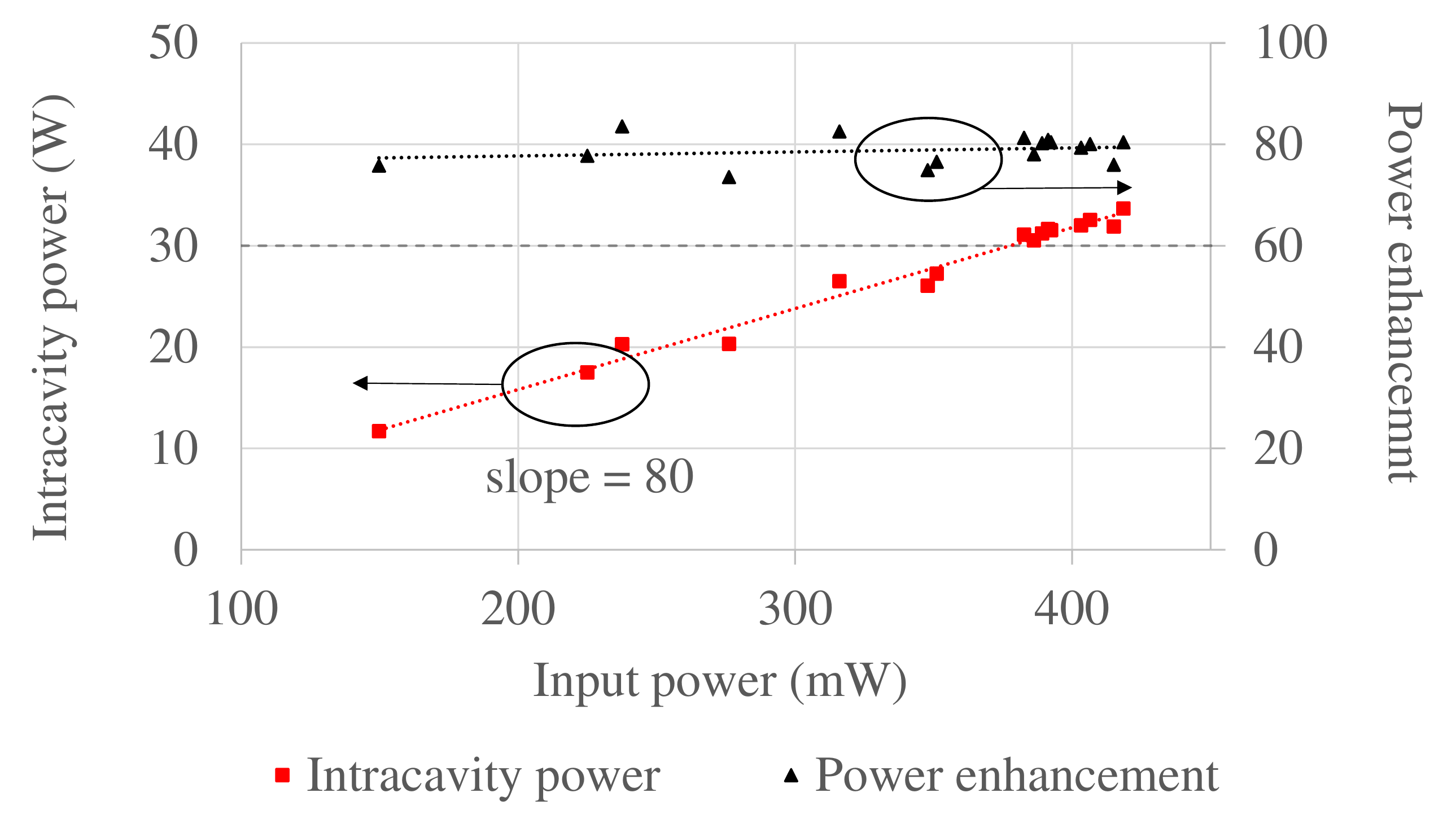}
\caption{Measured power enhancement factor and locked intracavity power as a function of input power. The maximum intracavity power is \SI{33.7}{\watt} with an input power of \SI{420}{\mW}. The maximum measured build-up factor was 84 and remained fairly constant over the range of input powers.}
\label{fig:results}
\end{figure}

We designed the enhancement cavity to operate within a differentially pumped vacuum system with a continuous inlet of ultra high purity oxygen gas. We tested the pressure limits for our mirrors at high locked intracavity powers (\numrange{10}{30} \si{\W}) and found that at oxygen pressures lower than \SI{500}{\milli\torr}, the mirrors would quickly degrade over the course of a few minutes as evidenced by severely reduced buildup factors. Inspection of the mirror surfaces after operation without an O$_2$ background revealed contamination on the surface which could be cleaned off with organic solvents.  Therefore, instead of oxygen depletion, we believe the main cause of degradation is due to residual hydrocarbons within the vacuum chamber which are cracked when exposed to high power UV radiation and contaminate the mirror surfaces \cite{Kunz:2000}. This supports the findings of \cite{Gangloff:15}, where they found a top surface layer of SiO$_2$ prevented degradation due to surface oxygen depletion. We observed that thorough cleaning of the vacuum chamber and optical components increased the run time before degradation. With \SI{500}{\milli\torr} of O$_2$ at the mirror surfaces, the optical cavity could be locked with $>$ \SI{30}{\W} of optical power for more than $\SI{1}{\hour}$ without evidence of degradation of the locked intracavity power and buildup. The linear dependence of the intracavity power as a function of input power shown in Fig. \ref{fig:results} suggests that the system is scalable to higher intracavity power.


\section{Discussion}
With the intracavity power achieved in this work, we are able to estimate a corresponding two-photon cooling rate. When the $2S_{1/2}$ state is fully mixed with the $2P_{3/2}$ state, the on-resonance scattering rate for the 1S-2S transition is $2.8~\times~10^{-7}~\si{\Hz~\,~\W^{-2}~\,~\cm^4} \, I_{UV}^2 $ \cite{kielpinski:2006}. For a reasonably sized beam waist of \SI{180}{\um}, \SI{30}{\W} of cw \uv radiation leads to a \SI{1}{\kHz} scattering rate. This scattering rate is modest by typical laser cooling standards, but when coupled with the high recoil velocity of hydrogen (\SI{3.3}{\m\per\s}) fairly rapid cooling cycles are possible.  For instance, the slow velocity tail ($<$ \SI{100}{\m\per\s}) of a collimated source of cryogenic hydrogen \cite{walraven:1982} could be cooled to the recoil velocity in \SI{30}{\ms}.

In addition to laser cooling for hydrogen and anti-hydrogen, this system could have applications in the spectroscopy of other two-body systems.  Notably, our system could be used for 1S-2S spectroscopy of muonium -- a bound state of an electron and an antimuon.  The most accurate measurement of this transition to date was performed in \cite{meyer:2000} using a pulsed laser with an \SI{86}{\ns} pulse duration.  In \cite{khaw:2016}, the spatial confinement of muonium was demonstrated and the possibility of using cw radiation to measure the 1S-2S transition was extensively discussed. When using cw radiation, systematics related to the finite pulse length (including  residual first-order Doppler shift, laser chirp, and AC Stark shifts) would be eliminated. The estimates presented in \cite{khaw:2016}, assumed 4 W of intracavity radiation which would result in a transit-time broadened linewidth of $\approx 1$ \si{\MHz}. With the increased DUV radiation demonstrated here, larger radiation beams could be used which would allow the recovered measurement linewidth to be reduced by a factor of $\approx \, 3$ if all other parameters are held constant.  Due to the difference in reduced mass, excitation of the 1S-2S transition in muonium requires radiation at \SI{244.2}{\nm}.  We confirmed that the system could be tuned to this wavelength and observed nearly identical performance of the enhancement cavity at the two wavelengths.

\section{Conclusion}

To conclude, we have demonstrated $> 30$ \si{\W} of power within a resonant enhancement cavity at \uv. To the best of our knowledge this is approximately ten times more than has been shown before \cite{beyer:2013}, and should enable a two-photon cooling scheme for atomic hydrogen.  At these high DUV intensities, a local pressure of O$_2$ is required to prevent mirror degradation.  Therefore, future cooling experiments will require additional stages of differential pumping to achieve a high-vacuum hydrogen beam line.  

\section{Acknowledgments}

We gratefully acknowledge useful discussions with Paolo Crivelli, Jacob Roberts, and Michael Morrison. Funding is provided from the NSF under Award \# 1654425.  

\bibliographystyle{ieeetr}
\bibliography{cavity-enhanced-duv}

\end{document}